\documentstyle[12pt]{article}

\newcommand{\nn}{\nonumber}
\newcommand{\vs}[1]{\vspace*{#1}}

\newcommand{\p}{\partial}
\newcommand{\Half}{\frac12}
\newcommand{\unit}{\hbox to 3.8pt{\hskip1.3pt \vrule height 7.4pt
    width .4pt \hskip.7pt \vrule height 7.85pt width .4pt \kern-2.4pt
    \hrulefill \kern-3pt \raise 3.7pt\hbox{\char'40}}}

\def\href#1#2{#2}

\newcommand{\alp}{{\alpha'}}
\newcommand{\calF}{{\cal F}}

\begin{document}

\begin{titlepage}

\bigskip

\title{
\vs{-10mm}
\hfill\parbox{4cm}{
{\normalsize\tt hep-th/0107226}\\[-5mm]
{\normalsize \tt NSF-ITP-01-74}
}
\\[50pt]
~\\
Non-Linear / Non-Commutative \\
Non-Abelian Monopoles
\\
~}
\author{
Koji {\sc Hashimoto}\thanks{\tt
koji@itp.ucsb.edu}
 \\ \\ [8pt]
{\it Institute for Theoretical Physics,
University of California, }\\
{\it Santa Barbara, CA 93106-4030}\\
}

\date{}

\maketitle
\thispagestyle{empty}

\begin{abstract}
\normalsize\noindent
Using recently proposed non-linearly realized supersymmetry in
non-Abelian gauge theory corrected to  ${\cal O} (\alp^2)$, we derive 
the non-linear BPS equations in the background $B$-field for the
$U(2)$ monopoles and instantons. We show that these non-Abelian
non-linear BPS equations coincide with the non-commutative
anti-self-dual equations via the Seiberg-Witten map.  
\end{abstract}

\end{titlepage}

It has been known that there are $\alp$  corrections to
the super Yang-Mills theory as a low energy effective action of
superstring theories \cite{gross}\cite{tsey}. The 
low energy effective theories have been a very strong tool for
analyzing the full string theory to find dualities and nonperturbative 
properties. However, the entire structure of the $\alp$ corrections
is still beyond our reach, although much elaborated work has been
devoted to this subject \cite{berg-o}--\cite{11}.
To be concrete, a stack of parallel D$p$-branes has a low energy
effective description which is the $p\!+\!1$-dimensional super
Yang-Mills 
theory accompanied with the $\alp$ corrections, but even in the
slowly-varying field approximation the complete form of the effective
action has not been obtained yet. To fix this problem, recently 
there appeared
several attempts to constrain the action by supersymmetries
\cite{ced}, or the equivalence \cite{SW} to non-commutative theories
\cite{col}\cite{teran}. Especially the paper \cite{ced} fixed all
the ambiguity of the ordering and coefficients up to 
${\cal O}(\alp^2)$.  

In this paper, we give an evidence supporting both of these
arguments of supersymmetries and non-commutative geometries, by
analyzing the BPS equations. 
Solitons and instantons as solutions of the BPS equations in the low
energy effective theory of D-branes have brane interpretations. For
example, a BPS monopole in U(2) Yang-Mills-Higgs theory corresponds to 
a D-string suspended between two 
parallel D3-branes. We consider these brane configurations in the
background $B$-field, and  explicitly construct $U(2)$ non-linear BPS
equations for the monopoles and the instantons. For the construction
we need an explicit form of the 
linearly/non-linearly realized supersymmetry transformations in the
effective theory which was obtained in \cite{berg-o} and \cite{ced}.  
According to the equivalence observed in \cite{SW}, these
equations should be equivalent with the $U(2)$ non-commutative BPS
equations \cite{AK}--\cite{gn}. In this paper we shall
explicitly show this equivalence\footnote{In the Abelian case, this
  equivalence was shown in \cite{HH}.}. This fact is a supporting
evidence of the supersymmetry transformation in the effective action
determined in \cite{ced}. Then we shall proceed to obtain the explicit 
solutions to these equations and discuss the brane interpretation of
them.  

The low energy effective action of open superstring theory with $U(N)$ 
Chan-Paton factor is given by the super Yang-Mills action corrected by 
$\alp$ \cite{gross}\cite{tsey}:
\begin{eqnarray}
{\cal L} = 
{\rm str}
\left[
-\frac14 (F_{ij})^2
+\Half \pi^2 \alpha'^2
\left(
  F_{ij}F_{jk}F_{kl}F_{li}- 
\frac14 (F_{ij})^2(F_{kl})^2
\right)
\right]
+ (\mbox{fermions})
+ {\cal O}(\alp^3).
\label{action}
\end{eqnarray}
The recent argument \cite{ced}\cite{berg-n} on the ordering of the
gauge fields and the fermions shows that up to the order of $\alp^2$
all the terms can be arranged by the symmetrized trace (str), which is 
compatible with the string scattering amplitudes and also the
supersymmetries. We use the action in the Euclidean four-dimensional
space to treat the anti-self-duality equation for both the monopoles
and instantons simultaneously. This action is obtained via 
dimensional reduction with $A_\mu = 0$ ($\mu = 0,5,6,7,8,9$).
The normalization for the
gauge symmetry generators is given by $  {\rm tr}[T^A T^B] =
\delta^{AB}$, which follows the convention of \cite{berg-n}. 

The action (\ref{action}) has a linearly realized supersymmetry
for the gaugino \cite{berg-o} 
\begin{eqnarray}
  \delta_\epsilon \chi^A
= \Half \Gamma_{ij} F^A_{ij}\epsilon
-\frac{1}{8}\pi^2\alp^2
{\rm str}(T^A T^B T^C T^D)
\left[
  F^B_{ij}F^C_{ji} F^D_{kl}\Gamma_{kl}
-4 F^B_{ij} F^C_{jk}F^D_{kl}\Gamma_{il}
\right]\epsilon,
\label{l}
\end{eqnarray}
which includes the $\alp$ corrections to the first nontrivial
order. The recent paper \cite{ced} shows that this system has another
supersymmetry, non-linearly 
realized supersymmetry, as is expected from the fact that the action 
(\ref{action}) describes a stuck of $N$ D-branes which breaks half of
the bulk 
supersymmetries. This non-linearly realized supersymmetry is given by 
\begin{eqnarray}
  \delta_\eta \chi^A = \eta^A
-\Half\pi^2(\alp)^2{\rm str} (T^A T^B T^C T^D)
\left[
  \Half \calF^B_{ij} \calF^C_{ij} 
+ \frac14 \calF^B_{ij}\calF^C_{kl}\Gamma_{ijkl}
\right]\eta^D,
\label{nl}
\end{eqnarray}
where the transformation parameter $\eta$ has its value only for a
$U(1)$ subgroup of $U(N)$ \cite{ced}. We have already neglected the
fermions in the right hand sides of (\ref{l}) and (\ref{nl}).

In order to compare our results with the previous
literatures \cite{HHM}--\cite{furu}\cite{HH}
we will consider only the gauge group $U(2)$. 
The normalized generators are defined as 
$  T^a = \frac1{\sqrt{2}}\sigma^a$ 
for $a = 1,2,3$ and $T^4 = \frac{1}{\sqrt{2}} \unit$.
Therefore especially the symmetrized trace of the four generators
appearing in the above supersymmetry transformations (\ref{l}) and
(\ref{nl}) is given by 
\begin{eqnarray}
  {\rm str}(T^A T^A T^A T^A) = 
  {\rm str}(T^a T^a T^4 T^4) = \Half,\quad\quad
  {\rm str}(T^a T^a T^b T^b) = \frac16\quad (a\neq b),
\end{eqnarray}
where the upper case $A$ runs all the generators of $U(2)$: $A =
1,2,3,4$. 

We turn on the background $B$-field which induces the
non-commutativity on the worldvolume of the D-branes. This $B$-field
is appearing in the action (\ref{action}) as $F^4_{ij} \rightarrow 
\calF^4_{ij} = F^4_{ij} +
2B_{ij}$, due to the bulk gauge invariance of the $B$-field. 

For simplicity, we put $\pi\alp = 1$, which can be restored on
the dimensional ground anytime. The action (\ref{action}) and its
symmetries (\ref{l}) (\ref{nl}) are obtained in string theory in the
approximation $\calF \ll 1$ and the slowly-varying
field approximation.   
We keep this in mind, and in the following we shall obtain the
non-linearly-modified BPS equations, 
perturbatively in small $B$. The basic BPS equations around whose
solutions we
expand the fields are the anti-self-duality equations 
\begin{eqnarray}
F_{ij}^{(0)a} + *F_{ij}^{(0)a}=0,
\quad F_{ij}^{(0)4}=0,
\label{asd}
\end{eqnarray}
where we have expanded the fields as $F_{ij}^A = F_{ij}^{(0)A} + {\cal
  O} (B)$, and the Hodge $*$ is defined as $*F_{ij}\equiv
\epsilon_{ijkl}F_{kl}/2$.
 These equations are obtained by considering the lowest order
in $\alp$ in (\ref{l}) by requiring a half of the linearly-realized
supersymmetries are preserved. The transformation parameters of the
preserved supersymemtries then obey the chirality condition 
\begin{eqnarray}
( 1+ \Gamma_5)\epsilon =0
\label{1+gamma5}
\end{eqnarray}
where $\Gamma_5 = \Gamma_{1234}.$
In the following, we assume that this chirality condition for
$\epsilon$ persists also to the 
higher order in $\alp$ and even with the inclusion of $B$. This
assumption will be checked by the explicit existence of the solutions. 

Along the argument given in \cite{SW}\cite{HH}--\cite{Moriyama2},
first we consider a combination   
of the two supersymmetries (\ref{l}) and (\ref{nl}) which remains
unbroken at the spatial infinity where $F=0$. The vanishing of $F$
gives 
\begin{eqnarray}
  \delta_{\epsilon} \chi^A = 
 B_{ij}\Gamma_{ij}\epsilon + {\cal O}(B^3), \quad
 \delta_{\eta} \chi^4 = 
\eta^4 + {\cal O}(B^2).
\end{eqnarray}
Thus 
$(\delta_{\epsilon} + \delta_{\eta}) \chi^4  = 0$ at the infinity is
equivalent with 
\begin{eqnarray}
\eta^4 = - B_{ij} \Gamma_{ij}\epsilon + {\cal O} (B^3).
\end{eqnarray}
Using this relation between two supersymmetry transformations, the
vanishing of the supersymmetry transform of the gaugino in all the
four-dimensional space leads to BPS conditions 
\begin{eqnarray}
&& \Half F_{ij}^a \Gamma_{ij}\epsilon =0,
\label{lbps}
\\
&& \Half \calF_{ij}^4 \Gamma_{ij}\epsilon 
- B_{ij}\Gamma_{ij}\epsilon
- \frac{1}{4}
{\rm str} (T^4 T^B T^C T^4)
\left[
  \Half \calF^B_{ij} \calF^C_{ij} 
+ \frac14 \calF^B_{ij}\calF^C_{kl}\Gamma_{ijkl}
\right]B_{ij}\Gamma_{ij}\epsilon
\nn\\
&& \hspace{35mm}- \frac{1}{8}
{\rm str} (T^4 T^B T^C T^D)
\left[
  \calF^B_{ij}\calF^C_{ji} \calF^D_{kl}\Gamma_{kl}
-4 \calF^B_{ij} \calF^C_{jk}\calF^D_{kl}\Gamma_{il}
\right]\epsilon
=0.
\label{com}
\end{eqnarray}
The first one (\ref{lbps}) gives usual anti-self-duality
equation\footnote{The $\alp$ corrections in the linearly-realized
  transformation (\ref{l}) are actually factored-out when 
  the lowest order relations (\ref{asd}) are substituted. }
without any correction of $B$.
In the analysis up to this order, 
only the $U(1)$ part of the gauge field obtains the first nontrivial
correction of $B$ as $  F_{ij}^{4} = {\cal O} (B)$.
Let us calculate the third and the fourth terms in (\ref{com}). 
Keeping in mind that we neglect the terms of the higher order, 
the third term can be arranged as
\begin{eqnarray}
- \frac{1}{4}
\left[
  (F^{(0)B}_{ij})^2
\right]B_{kl}\Gamma_{kl}\epsilon + {\cal O} (B^2),
\label{kekka}
\end{eqnarray}
where we have used the anti-self-duality of $F$ (\ref{asd}) and the
chirality of $\epsilon$ (\ref{1+gamma5}). 
After a straightforward calculation, the fourth term in (\ref{com})
can be evaluated in the same manner and turns out to be 
the same as (\ref{kekka})\footnote{These calculations are easily
  performed with the use of the block-diagonal form of the matrix $B$
  which is obtained by the space rotation without losing generality.}. 
The term $(F^{(0)})^3$ is negligible because it is of the higher
order. These evaluation simplifies the BPS condition (\ref{com}) to
\begin{eqnarray}
  F_{ij}^4 \Gamma_{ij}\epsilon -
(F^{(0)A}_{kl})^2 
B_{ij}\Gamma_{ij}\epsilon=0.
\end{eqnarray}
Decomposing this condition into the components, we obtain the {\it
  non-linear BPS equations}
\begin{eqnarray}
&&F_{ij}^4+ *F_{ij}^4 -\pi^2\alp^2 (B_{ij} + *B_{ij})
(F^{(0)A}_{kl})^2  =0,
\label{nlbps}
\end{eqnarray}
where we have restored the dimensionality.

The important is to check whether the equations (\ref{nlbps}) are
equivalent with the {\it non-commutative BPS equations}
via the Seiberg-Witten map \cite{SW}. 
The non-commutative U(2) monopoles/instantons \cite{AK}--\cite{gn} 
satisfy the
following BPS equations 
\begin{eqnarray}
  \hat{F}_{ij}^A + *\hat{F}_{kl}^A=0,
\label{ncbps}
\end{eqnarray}
where fields with the hat indicate the ones in the non-commutative
space. Substituting the Seiberg-Witten map \cite{SW}
\begin{eqnarray}
  \hat{F}_{ij} = F_{ij} + \Half \theta^{kl}
  \biggl(
    2\{F_{ik}, F_{jk}\}-\{A_k, (D_\l +\p_l) F_{ij}\}
  \biggr) + {\cal O}(\theta^2)
\label{sw}
\end{eqnarray}
into the above non-commutative BPS equation (\ref{ncbps}) 
and noting that the last gauge-variant terms in (\ref{sw}) 
vanish with the use of the lowest level anti-self-duality
(\ref{asd}), then we obtain 
\begin{eqnarray}
  F_{ij}^4 + *F_{ij}^4 + \frac14(\theta_{ij}+ *\theta_{ij})
(F_{kl}^{(0)})^2 =0.
\end{eqnarray}
Now we can use the relation \cite{SW, HH}
\begin{eqnarray}
  \theta_{ij} = -(2\pi\alp)^2 B_{ij}
\label{rela}
\end{eqnarray}
which has been  deduced from the worldsheet propagator for an open
string in the approximation $\alp B \ll 1$, 
then we can see the equivalence between the non-commutative BPS
equations (\ref{ncbps}) and the non-linear BPS equations
(\ref{nlbps}).   

Let us consider the specific brane configurations. 

\noindent
(1) \underline{$U(2)$ non-commutative monopole}.
In this case we perform the dimensional reduction further down to the
three-dimensional space and regard the fourth gauge field $A_4$ as a
scalar field $\Phi$. We turn on only one component of the
$B$-field, $B_{12}\neq 0$. Since we have a solution to the $U(2)$
non-commutative BPS equation for a monopole
\cite{HHM}\cite{Hata}\cite{gn}, and we know the Seiberg-Witten
transform of that solution to an appropreate order in $\alp$
\cite{HH}, then  
from the above equivalence, that transform is actually a corresponding 
solution to the non-linear BPS equation (\ref{nlbps}). 
After diagonalization of the scalar field, the eigenvalues 
exhibits the
configuration in which the single D-string suspended between the two
parallel D3-branes is tilted \cite{AK} so that they preserve 1/4
supersymmetries in the bulk with the $B$-field, as shown in \cite{HH}. 

\noindent
(2) \underline{$U(2)$ instanton}. It is known that the small instanton
singularity of the anti-self-dual instanton moduli space is resolved
if we introduce self-dual background $\theta$ \cite{SW, NS}. However,
this resolution does not occur in the case of anti-self-dual $\theta$.
This fact may be observed from the non-linear BPS equations and their
solutions. First let us analyze the anti-self-dual $B$-field (note the
relation (\ref{rela})) $B_{12}+B_{34}=0 $. Since the equation 
\begin{eqnarray}
  B_{ij}\Gamma_{ij} \epsilon=0
\end{eqnarray}
holds 
for $\epsilon$ which is involved with the preserved supersymmetries
for the anti-self-dual gauge field configuration, the whole $\eta$
terms vanish. Thus the linear BPS 
equation is not corrected, and so the configuration is not affected by
the $B$-field: 
\begin{eqnarray}
  F + *F=0.
\label{ord}
\end{eqnarray}
This is consistent with the observation that the linear BPS equation
$  \calF + *\calF=0 $
may solve fully $\alp$-corrected non-Abelian effective theory, as it
is true in the case of Abelian theory \cite{nocor}. Since now the 
self-duality is the same as the $B$-field orientation, we can subtract
the $B$-field from the both sides of the above equation and then
obtain (\ref{ord}).
This result may be related to the observation in \cite{HO} that
for the large instanton radius the commutative description of the
non-commutative $U(2)$ instanton \cite{furu} does not seem to have
$\theta$ dependence\footnote{For the small value of $\rho$ the gauge
  fields is not slowly-varying, the D-instanton
  charge distribution is corrected due to the derivative corrections
  to the Wess-Zumino term 
  \cite{wyllard}, thus we may not see any relation with \cite{HO}. }. 
From the non-commutative side, we substitute the Seiberg-Witten map to 
the non-commutative BPS equation (\ref{ncbps}), but then the order
$\theta$ terms cancel with 
each other and we found the usual anti-self-dual equation
(\ref{ord}). 

On the other hand, for the self-dual $B$-field background $B_{12} =
B_{34}$, there exists a
correction, which is expected from the resolution of the small
instanton singularity. 
One can solve the non-linear BPS equation (\ref{nlbps}) using the
general ansatz \cite{SW} in this background
\begin{eqnarray}
A_i^4 = B_{ij}  x^j h(r)
\end{eqnarray}
for a radial function $h(r)$.
Substituting the lowest order solution
\begin{eqnarray}
  F_{ij}^{(0)a} = \frac{4\rho^2}{(r^2+\rho^2)^2}\eta^{aij},
\end{eqnarray}
we obtain a differential equation for $h(r)$ and the solution is 
\begin{eqnarray}
  h(r) = 16\pi^2\frac{\rho^4(3r^2 + \rho^2)}{r^4 (r^2 + \rho^2)^3}.
\end{eqnarray}
This is the first nontrivial correction to the anti-self-dual
instanton.  
Since in this case the small instanton singularity must be resolved,
we might be able to see it by computing the instanton charge
distribution with this correction, but it turns out to be very small
as $\sim B^2 \rho^8 /r^{16}$
compared to the original instanton density 
$  \sim \rho^4/{r^8}$.
Therefore unfortunately we cannot see the change of the
instanton radius caused by the introduction of the $B$-field. 


\vs{10mm}
\noindent
{\bf Acknowledgments}: 
The author would like to thank T.\ Hirayama and W.\ Taylor for 
useful comments.  This research was supported in part
by Japan Society for the Promotion of
Science under the Postdoctoral Research Program (\# 02482),
and the National Science Foundation under Grant No.\ PHY99-07949.


\newcommand{\J}[4]{{\sl #1} {\bf #2} (#3) #4}
\newcommand{\andJ}[3]{{\bf #1} (#2) #3}
\newcommand{\AP}{Ann.\ Phys.\ (N.Y.)}
\newcommand{\MPL}{Mod.\ Phys.\ Lett.}
\newcommand{\NP}{Nucl.\ Phys.}
\newcommand{\PL}{Phys.\ Lett.}
\newcommand{\PR}{ Phys.\ Rev.}
\newcommand{\PRL}{Phys.\ Rev.\ Lett.}
\newcommand{\PTP}{Prog.\ Theor.\ Phys.}

\end{document}